\documentclass{emulateapj}
\usepackage{graphicx,rotating}

\def\be{\begin{equation}}
\def\ee{\end{equation}}
\def\ba{\begin{eqnarray}}
\def\ea{\end{eqnarray}}
\def\go{\mathrel{\raise.3ex\hbox{$>$}\mkern-14mu
             \lower0.6ex\hbox{$\sim$}}}
\def\lo{\mathrel{\raise.3ex\hbox{$<$}\mkern-14mu
             \lower0.6ex\hbox{$\sim$}}}

\def\bOmega{{\mbox{\boldmath $\Omega$}}}
\def\bu{{\bf u}}
\def\rmL{{\rm L}}
\def\kms{{\rm km~s}^{-1}}

\shorttitle{Mass Transfer, Transiting Stream and Magnetosphere in Exoplanets}
\shortauthors{Lai et al.}

\begin{document}

\title{Mass Transfer, Transiting Stream, and Magnetopause in Close-in 
Exoplanetary Systems with Applications to WASP-12}
\author{Dong Lai\altaffilmark{1,4}, 
Ch. Helling\altaffilmark{2,4},
and E.P.J. van den Heuvel\altaffilmark{3,4}}
\altaffiltext{1}{Department of Astronomy, Cornell University, Ithaca, NY 
14853, USA. Email: dong@astro.cornell.edu}
\altaffiltext{2}{SUPA, School of Physics and Astronomy, University of
St Andrews, North Haugh, St Andrews, KY16 9SS, UK}
\altaffiltext{3}{Astronomical Institute ``Anton Pannekoek'' and Center for 
High Energy Astrophysics, University of Amsterdam, The Netherlands}
\altaffiltext{4}{KITP, University of California, Santa Barbara, CA 93106}

\begin{abstract}

We study mass transfer by Roche lobe overflow in close-in exoplanetary
systems. The planet's atmospheric gas passes through the inner
Lagrangian point and flows along a narrow stream, accelerating to
100-$200~\kms$ velocity before forming an accretion disk. We show that
the cylinder-shaped accretion stream can have an area (projected in
the plane of the sky) comparable to that of the planet and a
significant optical depth to spectral line absorption. Such a
``transiting cylinder'' may produce an earlier ingress of the planet
transit, as suggested by recent HST observations of the WASP-12 system. 
The asymmetric disk produced by the accretion stream may also
lead to time-dependent obscuration of the starlight and apparent
earlier ingress.
We also consider the interaction of the stellar wind with the
planetary magnetosphere. Since the wind speed is subsonic/sub-Alfvenic
and comparable to the orbital velocity of the planet, the head of the
magnetopause lies eastward relative to the substellar line (the line
joining the planet and the star).  The gas around the magnetopause
may, if sufficiently compressed, give rise to asymmetric
ingress/egress during the planet transit, although more works are
needed to evaluate this possibility.

\end{abstract}


\section{Introduction}

The close-in exoplanets (with period less than a few days) discovered
in radial velocity and transit surveys are of great interest, not only
because they constrain theories of planet formation and evolution, but
also because they provide a probe of various physical processes that
are otherwise unimportant in ``normal'' planets.  WASP-12b is a
transiting exoplanet orbiting extremely close to a late-F/early-G
star ($M_\star=1.35M_\sun$, $R_\star=1.57R_\sun$, $T_{\rm eff}=6300$~K), 
with the orbital period $P=1.09$~days and orbital semi-major axis $a=0.023~{\rm
  AU}=4.94R_\sun =3.15R_\star$.  The planet mass $M_p=1.41M_J$ and
radius $R_p=1.79R_J$, as determined by transit observation in optical
continuum (Hebb et al.~2009; Campo et al.~2010). The planet is one of
the most irradiated exoplanets (with equilibrium temperature
$T_{\rm eq}=2500$-3000~K) and is highly inflated. The fact that
the Roche radius (Hill sphere radius) of the planet,
$R_L=a(M_p/3M_\star)^{1/3}=1.85R_p$, is only slightly larger than
the $R_p$ derived from optical transit measurements,
suggests that mass loss from the planet is likely (Li et
al.~2010). The small orbital separation also suggests that stellar
wind may influence the atmosphere the planet.

Recently, Fossati et al.~(2010) obtained near-UV transmission
spectroscopy of WASP-12b with the Cosmic Origins Spectrograph on the
{\it Hubble Space Telescope}. The data revealed enhanced transit
depths (by about a factor of 2) in two wavelength bands, NUVA
(2539-2580~$\AA$) and NUVC (2770-2811~$\AA$), which were attributed
to flux attenuation by absorption lines of metals in the vicinity of the
planet.
Most interestingly, the NUVA data exhibits an earlier ingress
compared to the transit in optical J,B and Z bands, while the egress
of the transit occurs at about the same time as the optical transit.

The asymmetric behavior of the ingress/egress in the NUVA band
relative to the continuum is difficult to understand if the absorbing
gas surrounding the planet arises entirely from an irradiation-driven wind
-- such a wind has been studied extensively in the context of hot Jupiter 
HD 209458b (e.g., Yelle 2004; Tian et al.~2005; Garcia Munoz 2007; 
Murray-Clay et al.~2009):
The wind is most strongly generated at the planet's dayside,
and would be distributed on both sides (terminations)
of the substellar line (the line joining the planet and the star). 
If anything, it would be 
preferentially on the west side because of
the planetary rotation, which is almost likely synchronized with
the orbit (see Schneiter et al.~2007 for a simulation).

In this paper, we consider two possible explanations for the
asymmetric excess absorption during WASP-12b transit. First, we study
the Roche lobe overflow from the planet to the parent star (see Li et
al.~2010) and the associated accretion stream. The stream is
asymmetric with respect to the line joining the planet and the star. We
show that the stream has a sky-projected area comparable to the
projected planet area, and a sufficiently large column density to
cause blockage of the star light prior to optical ingress. Second, we
qualitatively discuss the magnetopause produced by the interaction between the
stellar wind and the planet's magnetosphere. Because the planet's
orbital velocity, $v_{\rm orb}=(GM_\star/a)^{1/2} =228~\kms$, is
comparable to the stellar wind velocity ($v_w\simeq 100~{\rm
  km~s}^{-1}$ at $a=0.023$~AU), the head of the magnetopause lies
eastward relative to the substellar point. Again, 
ingress/egress asymmetry may be produced if the gas density in 
around the magnetopause is sufficiently high.


This paper is organized as follows. In Sect.~2 we discuss line 
absorption by a moving medium with a velocity gradient and derive
the observational constraints on the absorbing gas in WASP-12b. 
In Sect.~3 we study the property of the accretion stream 
and the obscuration of the star light by the stream.
Section 4 examines the possibility that the gas around the magnetopause 
may absorb the star light. We conclude in Sect.~5.

\section{Observational Constraint on the Excess Absorbing Gas}

We first consider line absorption of stellar radiation by the gas
around the planet and beyond (e.g., in the accretion stream).  The
absorbing gas has a finite temperature $T$ and line-of-sight bulk
velocity $V_\parallel$, both of which (in general) depend on the
spatial position ${\bf r}$. We denote the velocity distribution of the
column density of the gas (of a given species i) by
\be 
{dN_i\over dV_\parallel}=\int\! dl
{dn_i(V_\parallel,l)\over dV_\parallel},
\ee 
where the integration is along the line of sight, and
$dn_i/dV_\parallel$ is the velocity distribution of the gas density.
The stellar radiation intensity $I_\nu^{(0)}$ is attenuated to
$I_\nu=I_\nu^{(0)}\exp(-\tau_\nu)$ after passing through the absorbing gas,
with the optical depth given by 
\be
\tau_\nu=\int\!\sigma_\nu(V_\parallel)\,{dN_i\over dV_\parallel}dV_\parallel.
\ee
The line absorption cross section can be written as (e.g., 
Rybicki \& Lightman 1979)
\be
\sigma_\nu(V_\parallel)=f{\pi e^2\over m_ec}\Phi[\nu-\nu_0(1+V_\parallel/c)],
\ee
where $f$ is the oscillator strength of a specific line transition, and
$\Phi$ is the Voigt profile centered at $\nu_0(1+V_\parallel/c)$ (with
$\nu_0$ the intrinsic line frequency). When thermal broadening 
dominates, we have $\Phi=1/(\sqrt{\pi}\Delta\nu_T)$ at the line center, 
where $\Delta\nu_T$ is the thermal width. Since $\Delta\nu_T$ is much 
less than the bandwidth of transiting observations, the Voigt profile 
can be approximated by a delta function. Thus we have
\be
\tau_\nu=f{\pi e^2\over m_ec\nu_0} c\left({dN_i\over dV_\parallel}\right)_{V_\parallel
=c(\nu-\nu_0)/\nu_0},
\ee
where $(\pi e^2/m_e c\nu_0)=(10^{-16}/4)(\lambda_0/2800\,\AA)\,{\rm cm}^2$.

For WASP-12b, Fossati et al.~(2010) found that the transit depths in
the wavelength regions NUVA (2539-2580~$\AA$) and NUVC
(2770-2811~$\AA$) are larger than the continuum depth by a factor of
two at the 2.5$\sigma$ level, implying the effective radii of order
$2.69R_J$ and $2.66R_J$, respectively. The flux depletions in the
41~$\AA$ window of NUVA and NUVC are produced by many absorption
lines. If we require a single line (of a given atomic species i) to
significantly deplete the flux (i.e., $\tau_\nu\go 1$) in the
wavelength range $\Delta\lambda$ around its intrinsic wavelength
$\lambda_0$, then the column density of the gas with velocity around
$V_\parallel=c\Delta\lambda/\lambda_0$ must satisfy
\ba
&& \left({dN_i\over d\ln
  V_\parallel}\right)_{V_\parallel=c\Delta\lambda/\lambda_0}\!\!\!\!\!\go 
{4\times 10^{13}\over f}
\!\left(\!{2800\,\AA\over\lambda_0}\!\right)\!\!
\left(\!{\Delta\lambda/\lambda_0\over 10^{-3}}
\!\right)\,{\rm cm}^{-2}\nonumber\\
&&\qquad \simeq  {1.3\times 10^{13}\over f}\!
\!\left(\!{2800\,\AA\over\lambda_0}\!\right)\!\!
\left(\!{V_\parallel\over 100\,\kms}\!\right)\,{\rm cm}^{-2}.
\label{eq:Ni}\ea
For example, for the Mg~II~2800~$\AA$ lines, $f=0.93$, and if we take
$\Delta\lambda =1~\AA$, we find
$\left({dN_i/d\ln V_\parallel}\right)\go 1.4\times 10^{13}~{\rm cm}^{-2}$
at $V_\parallel\simeq 108~\kms$.
For a characteristic length scale of $3R_p=5.4R_J$ (see below), this corresponds
to the gas density $dn_i/d\ln V_\parallel\go 400~{\rm cm}^{-3}$.
Note that, since the thermal velocity of the Mg atom is $v_T=(kT/24m_H)^{1/2}=1.9\,
(T/10^4\,{\rm K})^{1/2}~{\rm km~s}^{-1}$, 
such a large $\Delta\lambda/\lambda_0$ ($=1/2800$, corresponding to
$V_\parallel=c\Delta\lambda/\lambda_0=108~\kms$)
cannot be due to thermal broadening, but must arise from significant bulk motion
of the absorbing gas along the line of sight. If we choose a smaller $\Delta\lambda$,
(i.e., a single line covers a smaller range of wavelength in the absorption),
the required column density is smaller, but then we would require
a denser spectrum of absorption lines so that the whole $41~\AA$ wavelength
regions of NUVA and NUVC are significantly absorbed.

Now consider the ingress/egress asymmetry in the transit of the NUVA
band.  Figure 2 of Fossati et al.~(2010) shows that in optical
continuum, the ingress starts at the orbital phase $\phi\simeq 0.945$,
and the center of the transit is at $\phi=1$. The phase difference of
0.055 corresponds to a transverse distance (projected on the plane of
the sky) of $0.925 R_\star+R_p\simeq 1.04R_\star$ (for the orbital
inclination angle $i=83.1^\circ$).  In the NUVA band, the ingress starts
at $\phi\simeq 0.925$ or earlier. The phase difference of 0.075 then
corresponds to a transverse distance of
$(0.075/0.055)(1.04R_\star)=1.42R_\star$. 
This implies that the excess NUVA absorber during the
ingress must extend a distance of at least 
$0.38R_\star\simeq 3.2R_p$ from the planetary surface.

The above constraints for the column density and size/location of the
line absorber will be used to evaluate two possible models for the
ingress/egress asymmetry in the following sections..

\section{Roche Lobe Overflow and Accretion Stream}

\subsection{Accretion Stream and Disk Formation}

The gas dynamics of Roche lobe overflow in semidetached binaries
was studied by Lubow \& Shu (1975,1976).
Here we adapt their theory to mass transfer in close-in exoplanetary
systems like WASP-12b (cf.~Li et al.~2010).

Consider the atmosphere of the planet in quasi-hydrostatic equilibrium
extending to its Roche lobe.
As the gas flows through 
the L1 point, it transitions from subsonic to supersonic, with the
sonic surface close to the L1 point. The size of the
transonic nozzle at L1 is 
\be
\Delta R_\rmL\sim c_s/\Omega_o,
\ee
where $c_s$ is the sound speed and $\Omega_o=2\pi/P$ is orbital
frequency of the binary. 
The mass transfer rate is related to the gas density at L1 by
\be
\dot M\sim \pi\rho_\rmL c_s (\Delta R_\rmL)^2.
\ee

The steady-state equation of motion of the gas in the rotating frame reads
\be
{d\bu\over dt}=(\bu\cdot\nabla)\bu=-\nabla\Phi-2\bOmega\times\bu-{\nabla P\over
\rho},
\ee
where $P=\rho c_s^2$ is the gas pressure, and $\Phi$ is the
Roche potential:
\be
\Phi=-{GM_\star\over r_1}-{GM_p\over r_2}-{1\over 2}\Omega_o^2
\left[(X+a-\mu a)^2+Y^2\right].
\ee
Here $\mu=M_\star/(M_\star+M_p)$, $r_1=r=\sqrt{X^2+Y^2}$,
$r_2=\sqrt{(X+a)^2+Y^2}$, and the coordinates $(X,Y)$ are measured from the center
of the star (see Fig.~1). The L1 point is determined from 
$\partial\Phi/\partial X=0$ for $Y=0$, giving the distance of L1 from 
the center of the planet:
\be
x_L=a-|X_L|\simeq a\left({q\over 3}\right)^{1/3}\left[1-{1\over 3}
\left({q\over 3}\right)^{1/3}\right],
\ee
for $q=M_p/M_\star\ll 1$. The gas along a streamline conserves its
Bernoulli integral (assuming constant $c_s$),
${\cal B}=\bu^2/2+\Phi+c_s^2\ln\rho$.
As the gas leaves L1 and accelerates toward the star, pressure becomes
negligible at a distance larger than $\Delta R_\rmL$, and then the gas particle
follows a ballistic trajectory. Due to the combined effects of Coriolis force and
Roche potential force, the gas particle leaves the L1 region in a narrow
stream (with width $\Delta R_\rmL$), along the direction given by the angle
$\theta_s$ (the angle between streamline near L1 and the X-axis)
\be
\cos 2\theta_s=-{4\over 3A}+\left(1-{8\over 9A}\right)^{1/2},
\ee
with 
$A=a^3\left[{\mu/r_1^3}+{(1-\mu)/r_2^3}\right]_{\rm L1}$.
Thus $\theta_s=26.7^\circ$ for $q=10^{-3}$ (and $28.4^\circ$ for $q\rightarrow 0$).
Figure 1 illustrates (in the case of $q=10^{-3}$) the stream trajectory
for several values of initial gas velocities comparable to the sound speed
[$c_s/(a\Omega_o)=0.02$, or $c_s\simeq 5~{\rm km~s}^{-1}$ for the WASP-12 system.]

%
\begin{figure}
        \vskip -0.7truecm
        \includegraphics[width=0.51\textwidth]{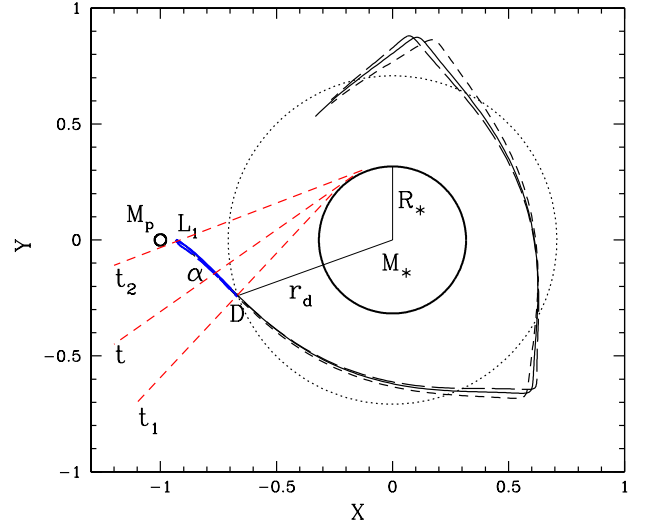}
        \vskip -0.99truecm
        \caption{
Stream trajectory and disk formation in the orbital plane for the case of
planet-star mass ratio $q=M_p/M_\star=10^{-3}$ (WASP-12b
parameter). The planet is located at $(X,Y)=(-1,0)$ (in units of $a$),
and rotates counter-clockwise around the star.
Three streamlines leaving L1 are shown, corresponding
to the initial velocity (in units of the orbital velocity of the planet,
$\sqrt{GM_\star/a}=228~{\rm km~s}^{-1}$) 
of $(u_x,u_y)=(0,0)$ (solid line),
$(0.02,0)$ (dashed line) and $(0.02,-0.02)$ (long-dashed line).
In the absence of a disk, matter would follow these stream
trajectories. Self-collisions of the streams cause
the formation of a disk.
The disk outer radius is at $r_d=0.71$. The stream strikes the disk
at point D, beyond which the stream does not exist. The dashed straight line 
labeled $t_1$ indicates the line of sight when the star light is first
blocked by the stream (the ingress of the ``stream transit''), and the
line labeled $t_2$ indicates the ingress of the ``normal'' planet transit.
A general line of sight is labeled $t$, with the angle between the
stream axis and the line of sight denoted by $\alpha$.}
\end{figure}

The crossings of streamlines lead to shock dissipation and disk formation.
The stream collides with the disk at point D (see Fig.~1), whose position
(and thus the outer radius of the disk $r_d$) is determined by the
``no-slip'' condition, i.e., 
the $\phi$-component of the incident stream
velocity equals the circular velocity at the same point, 
the latter being given by (in the corotating frame)
$v_{\rm circ}\simeq (GM/r)^{1/2}-r\Omega_o$ (for $q\ll 1$).
Figure 2 shows the velocity of the accretion stream before it collides
with the outer edge of the disk for the cases of $q=0.001$ and $0.005$.
We see that, for example, for $q=0.001$, the stream
impacts the disk at $|Y_d|=0.24a$ and $r_d=0.71a$, with the 
terminal velocity $u=0.53a\Omega_o$ (and 
$u_x$ reaches $0.38a\Omega_o$).

\begin{figure}
        \vskip -0.8truecm
        \includegraphics[width=0.51\textwidth]{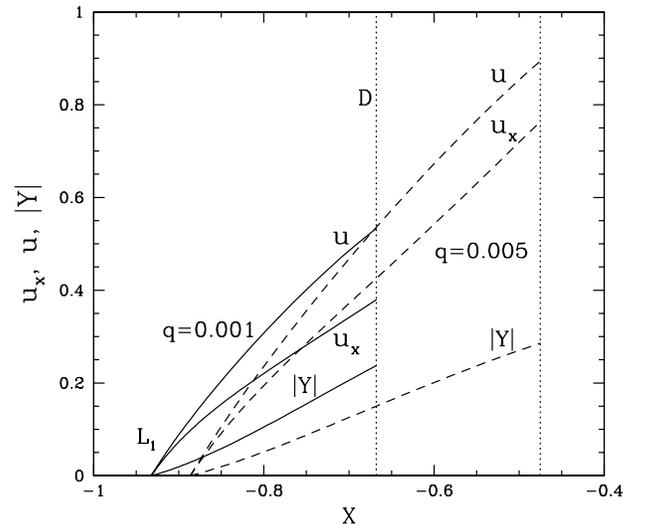}
        \vskip -1.2truecm
        \caption{
Stream trajectory and velocity for mass ratio
$q=0.001$ (solid lines) and $0.005$ (dashed lines). The stream
starts at L1 and terminates when it collides with the outer edge of the 
disk (the vertical dotted line labeled ``D''). The star and planet are 
located at $(X,Y)=(0,0)$
and $(-1,0)$ (in units of $a$), respectively. The plotted quantities are
$u$ (stream velocity in the rotating frame, 
in units of the orbital velocity of the planet,
$\Omega_o a=228~{\rm km~s}^{-1}$ for WASP 12-b), 
$u_x$ (the $X$-component of the stream velocity) and $|Y|$ (the $Y$-position
of the stream; see Fig.~1).}
\end{figure}


As the stream evolves from L1 to impact (point D), its width $W$ 
undergoes relatively small variation, due to the combined effects of
the finite sound speed (which tends to make the stream expand) and 
the enhanced gravity closer to the star (which tends to compress the
stream). Thus, $W\simeq \Delta R_L$, and the
cross-sectional area of the stream remains of order $\pi(\Delta R_L)^2$.
The mean density of the stream therefore varies as 
\be
\rho_s\sim \rho_L {c_s\over u},
\ee
for the stream velocity $u\go c_s$.

Note that the ``no-slip'' condition likely only gives the minimum value of
the outer disk radius (see Shu \& Lubow 1981). 
In reality, tidal forces from the planet set
the outer truncation of the disk.  Numerical simulations of the
planet-disk interaction in the Type II regime (in which the planet is
sufficiently massive to open a gap in the gas disk; see Lin \& Papaloizou
1993) typically find that the gap half width is about 0.2-$0.3a$ (e.g.,
Lubow \& D'Angelo 2006; Armitage 2007). Thus we shall
adopt the outer disk radius $r_d\simeq 0.7a$ in this paper.


The accretion disk formed from the stream has a steady-state surface density 
$\Sigma_d(r)\simeq \dot M/(3\pi\nu_d)$, where $\nu_d$ is the disk viscosity.
Adopting the $\alpha$-viscosity ansatz, $\nu_d=\alpha_d H_d c_{ds}$,
where $c_{ds}$ is the disk sound speed and $H_d=c_{ds}/\Omega_d(r)$ is the
disk thickness, we find 
\be
\Sigma_d\simeq {1\over 3\alpha_d}\left({c_s\over c_{ds}}\right)^2\left({\Omega_d\over
\Omega_o}\right)\Sigma_L,
\label{eq:sigmad}\ee
where $\Sigma_L\simeq\rho_L W$ is the gas column density at the L1 point.
Obviously, for $\alpha_d\ll 1$, we expect $\Sigma_d\gg\Sigma_L$.

\subsection{Application to the WASP-12 System}

To estimate the gas density of the accretion stream in the WASP-12 system,
we assume that the planet has a isothermal atmosphere with temperature
$T\simeq 3000$~K (corresponding to a sound speed of $c_s\simeq
5$~km~s$^{-1}$)\footnote{The equilibrium temperature of the
planet ranges from $2500$~K to 3000~K, depending on the albedo and
the efficiency
of heat redistribution in the planetary atmosphere.}
and surface density $\rho(R_p)\sim 5\times
10^{-8}~{\rm g~cm}^{-3}$ (see Li et al.~2010).
Relative to the planet center, the Roche lobe extends to $L1$ at $x_L\simeq
a(q/3)^{1/3}=1.85~R_p$. Along the $y$-direction (in the orbital plane),
it extends to $y_L=(2/3)x_L=1.23R_p$. Since the Roche potential along the
planet's $y$-axis (relative to the planet center) is simply $-Gm/y+$constant, 
one finds $\rho(y_L)\sim 3\times 10^{-12}~{\rm g~cm}^{-3}$ (Li et al.~2010).
Since the Roche lobe is an equipotential surface, this is also close to
the density at L1. With the nozzle radius $\Delta R_L\sim 7.5\times 10^4~{\rm km}
\simeq 0.6R_p$ and a thermal outflow velocity $c_s$, 
we find that the mass transfer rate through $L1$ is
$\dot M\sim 2.7\times 10^{14}~{\rm g~s}^{-1}\simeq 3\times 10^{-9}
M_p~{\rm yr}^{-1}$.

Note that Li et al.~(2010) considered mass transfer driven by an
energy source within the planet's Roche lobe. Such an energy source
may lie inside the planet, maintaining its inflated radius. The
suggested energy source was tidal dissipation due to an assumed 
eccentricity of the orbit (see also Gu et al.~2003). It has
since, however, become clear that the orbit of WASP-12b is highly circular
($e=0.017\pm^{0.015}_{0.011}$ from new radial-velocity monitoring;
Husnoo et al.~2010), implying that such an eccentricity-driven tidal
energy source is negligible. Li et al.
obtained a much higher mass loss rate (about 25 times larger than our
estimate) as they assumed that the matter flows from below the
Roche lobe over the entire surface area of the Roche lobe. 
In our discussion below, we take a conservative approach and 
adopt our (smaller) value of $\dot M$ given above,
as we believe it sets the minimum value of the mass transfer rate.

The above discussion neglects the possible effect of the planet's
thermosphere, where photo-ionization heats the gas temperature to $\sim
10^4$~K (e.g., Murray-Clay et al.~2009; Koskinen et al.~2010).
Taking the photo-ionization cross section $\sigma_{\rm PI}\simeq
6\times 10^{-18}(\epsilon_0/13.6~{\rm eV})^{-3}$~cm$^2$ ($\simeq
2\times 10^{-18}$~cm$^2$ for photon energy $\epsilon_0=20$~eV), we find that the
column density that UV photons can penetrate is $5\times
10^{17}$~cm$^{-2}$. For $\rho_L\sim 3\times 10^{-12}$~g~cm$^{-3}$, this
corresponds to a distance of 2.8~km, much less that $\Delta R_L$. Thus, only the
``skin'' of the accretion stream will be affected by photo-ionization.
This also shows that, the planet's thermosphere, if any, lies outside the 
planet's Roche lobe.

In principle, the Roche lobe, the accretion stream and the disk can all block
the star light. Here we focus on the accretion stream since it is asymmetric 
relative to the line joining the star and the planet.
To estimate the transiting property of the stream, we model it
as a cylinder with length $s_d$ and radius $W=\Delta R_L$.
The result of Sect.~3.1 gives $s_d=0.36a$ (for $X_d=-0.668$ and $Y_d=-0.238$;
see Fig.~2).
Thus the area of the stream as projected in the plane of the sky
is $s_d W\sin\alpha\simeq 6R_p^2\sin\alpha$, where $\alpha$
is the angle between the line of sight and the symmetry axis of the 
cylinder. Note that $\alpha$ varies during the transit (see Fig.~1). Clearly,
if the stream is opaque to the radiation, it would lead to significant 
blockage of the star light, comparable to that produced by the planet itself.

To evaluate the column density of the stream, we need to know how the
density varies along the stream. As discussed in Sect.~3.1, in the
supersonic region, the density decreases as $u^{-1}$ as the stream
velocity $u$ increases. 
Figure 2 shows that $u$ increases approximately as a linear function
of $s$, the distance from L1 along the stream. 
Thus we adopt $u(s)\simeq c_s/F(s)$, with $F(s)=1$ for $0<s<W$ and $F(s)=W/s$ for
$W\le s<s_d$. This gives $\rho\simeq \rho_L F(s)$.
The hydrogen column density in the stream along the line of sight is
then $N_H(s)\simeq (\rho/m_H)(W/\sin\alpha)\sim 10^{22}
F(s)/\sin\alpha$~cm$^{-2}$. With $V_\parallel=c_s (s/W)\cos\alpha$ for
$s>W$, we find
\be
N_H\sim 10^{21}\left({100\,\kms\over V_\parallel\tan\alpha}\right)
\,{\rm cm}^{-2}
\ee
for $V_\parallel\go 5\cos\alpha\,\kms$ 
(where we have used $c_s=5\,\kms$). 
Thus, as long as the metal abundance satisfies
${N_i/N_H}\go 10^{-8}f^{-1}\tan\alpha
\left({V_\parallel/100\,\kms}\right)^2$,
the stream column density would be larger than that
required by Eq.~(\ref{eq:Ni}). For example, 
the solar abundance of Mg relative to H is $3\times 10^{-5}$,
and the Mg~II~2800~$\AA$ lines (with $f=0.93$) may then produce
significant absorption of the star light.
For Scandium, the solar abundance is about $10^{-9}$, thus stronger
lines with $f\go 10$ or a super solar abundance would be needed to
produce significant line absorption
(note that WASP-12a is supersolar by about 0.3; Hebb et al.~2009).
We note that if the gas velocity (at a given $s$) 
is uniform within the cross-section of the stream, then the difference
in the gas velocity along a given line-of-sight is 
$\Delta V_\parallel\sim c_s\cos^2\alpha/\sin\alpha$. This is too small to
produce significant reduction in the observed flux. However, it is 
possible that the 3D gas velocity structure in the stream 
is more complicated (e.g., due to the turbulence developed at 
the surface of the stream via Kelvin-Helmholtz instability) 
and a larger velocity difference along the line-of-sight could be produced.

The collision between the accretion stream and the disk produces a hot
spot, which may lead to time-dependent obscuration of the star light
and possibly early ingress (C. Baruteau \& D. Lin, 2010, private
communication).  As shown above [see Eq.~(\ref{eq:sigmad})], the
accretion disk formed from the stream has a significant surface
density, and can cause obscuration of the star light. Obviously, a
perfectly symmetric disk would only give time-independent obscuration.
However, disk asymmetry (e.g., produced by the hot spot) may cause an
apparent earlier ingress of the planet transit. Numerical simulations
will be needed to assess these possibilities.

Our discussion so far has focused on the accretion stream due to mass
transfer through the inner Lagrangian point L1. In general, an
outward-spiraling stream may also form due to mass loss through the L2
Lagrangian point (S. Lubow, private communication)
\footnote{G. Trammell, P. Arras and Z. Li (2010, to be submitted)
have also recently considered this issue.}. The L2 point is located at a distance
$x_{L2}\simeq a\left({q/3}\right)^{1/3}\left[1+(1/3)
\left({q/3}\right)^{1/3}\right]$ from the center of the planet,
farther away from the host star than the planet. The (Roche) potential difference 
between L1 and L2 is 
\be
\Phi_{\rm L2}-\Phi_{\rm L1}={2\over 3}{GM_p\over a}.
\ee
If we assume that the Roche lobe overflow through L1 does not affect
the density of the outer L2 Roche lobe, we can estimate the
density at L2 as $\rho_{\rm L2}\simeq \rho_{\rm L1}\exp(-2GM_p/3ac_s^2)$.
For the WASP-12b parameters, we find $\rho_{\rm L2}/\rho_{\rm L1}\simeq 0.2$
for $c_s=5\,\kms$ (see also Gu et al.~2003). In reality, 
$\rho_{\rm L2}/\rho_{\rm L1}$ will be somewhat smaller than the above 
estimate, and a lower night-side temperature of the planet 
will also reduce $\rho_{\rm L2}$. We conclude that 
the spiral stream from L2 is less important than the accretion stream. 
Numerical simulations of the two streams would be useful to accurately evaluate
their relative effects on the planet transit.

\section{Magnetopause}

The star WASP-12a resembles the Sun in many aspects,
and we will parametrize its wind property using the observed
solar wind parameters. The fiducial solar wind mass flux is 
$\dot M_w=3\times 10^{-14}M_\sun\,{\rm yr}^{-1}$.
At 1~AU, the typical solar wind speed is of order $450~\kms$
(but can change on various timescales from $200~\kms$ to $1000~\kms$),
and the electron density, temperature and magnetic field in wind are 
$\sim 7$~cm$^{-3}$, $1.4\times 10^5$~K, and $7\times 10^{-5}$~G, respectively
(e.g., Hundhausen 1995). Thus, the solar wind at 1~AU has a plasma $\beta$ 
(the thermal pressure divided by the magnetic pressure) of order unity, and the fast
magnetosonic Mach number is about 6 (Russell \& Walker 1995).


To obtain the stellar wind property at the distance appropriate to
WASP-12b ($a=0.023$~AU) requires detailed wind models, which we will
not attempt here. But it is very likely that at such a small
distance, the stellar wind is subsonic and sub-Alfvenic, and the fast
magnetosonic Mach number is less than unity (e.g., Mestel \& Spruit 1987;
Cranmer 1998). For example, using the
simplest (isothermal) Parker wind solution (for $M_\star=1.35M_\sun$),
we find that if the wind speed at 1~AU is $500~\kms$, then the sonic
point is at $0.032$~AU, and the wind velocity and Mach number at
0.023~AU are $90~\kms$ and 0.66, respectively. (For a wind speed of
$600~\kms$ at 1~AU, the corresponding numbers become 0.024~AU, $148~\kms$
and 0.94.)  The solar wind temperature actually increases with
decreasing distance, and this will make the Mach number at 0.023~AU
even smaller.  The plasma $\beta$ in the wind may reach below 0.1 at
such a distance, with the wind magnetic field dominated by the radial
component.


Given that at 0.023~AU, the stellar wind has a fast magnetosonic Mach
number less than unity, we do not expect bow shock to form when the
wind interacts with the planet's magnetic field. Instead, near the
planet, the wind will be ``gradually'' stopped at the magnetopause.
Importantly, since the orbital velocity of the planet, $v_{\rm
  orb}=228~\kms$, is comparable to the wind velocity at $0.023$~AU,
$v_w\sim 100~\kms$, the head of the magnetopause is shifted eastward
with respect to the substellar point by an angle $\tan^{-1}(v_{\rm
  orb}/v_w)$ ($=66^\circ$ for $v_w=100~\kms$).

In general, the location of the magnetopause
is determined by balancing the momentum flux $(P+\rho v_\perp^2
+B_\parallel^2/8\pi)$
across the contact surface, where $P$ is the gas pressure, $v_\perp$ is the
gas velocity perpendicular to the surface, and $B_\parallel$ is
the magnetic field
parallel to the surface (e.g., Shu 1992).
The standoff distance $r_m$ (measured from the center of the planet)
at the head of the magnetopause is obtained by 
\be
\rho_w (v_w^2+v_{\rm orb}^2)+P_w+{B_{w\parallel}^2\over 8\pi}
=P_p+{B_p^2\over 8\pi}\left({R_p\over r_m}\right)^6,
\label{eq:balance}\ee
where $P_p$ is the gas pressure of the planetary magnetosphere, and
we have assumed that the planet has a dipolar magnetic field
with the equatorial strength $B_p$. 
In the absence of a detailed wind model, we parameterize
Eq.~(\ref{eq:balance}) by
\be
f_w\rho_w v_w^2=f_p
{B_p^2\over 8\pi}\left({R_p\over r_m}\right)^6.
\ee
We estimate $f_w\simeq 1+v_{\rm orb}^2/v_w^2+{\cal M}^{-2}\sim 10$
(for $v_w\sim 100~\kms$ and the fast magnetosonic Mach number 
${\cal M}\sim 0.5$), and $f_p\sim 1$. With 
$\dot M_w=4\pi a^2\rho_w v_w$, we then have
\be
{r_m\over R_p}=2.6\,f_p^{1/6}\!
\left(f_{w1}v_{w1}M_{w1}\right)^{-1/6}\!
B_{p1}^{1/3},
\label{eq:rm}\ee
where $f_{w1}\equiv f_w/10$, $v_{w1}\equiv v_w/(100\,\kms)$,
${\dot M}_{w1}\equiv \dot M_w/(3\times 10^{-14}M_\sun\,{\rm yr}^{-1})$
and $B_{p1}=B_p/(1\,{\rm G})$.
We see that the main uncertainty in our estimate of $r_m$ is the
magnetic field strength of the planet. For example, if $B_p\sim 5$~G,
the magnetopause may extends to $4.4R_p$.
Note that $r_m$ is similar to the required extension of the absorbing
gas to produce the asymmetric ingress/egress in WASP-12b (see Sect.~2).
This motivates us to consider the possibility that the gas 
(from the stellar wind) around $r_m$ may absorb the star light.

The number density of H in the stellar wind (before any compression)
at $a=0.023$~AU is given by 
\be
n_H^{(0)}={\dot M_w\over 4\pi a^2v_wm_H}=7.7\times 10^4\,
{\dot M}_{w1}v_{w1}^{-1}\,{\rm cm}^{-3}.
\label{eq:nH0}\ee
As the stellar wind gas is deflected at the magnetopause, it may
undergo some compression (of order unity) and move around the field
lines at the velocity comparable to $v_w\sim 100~\kms$.
Whether this gas can provide enough line absorption
to explain the observation depends on the physical states of the gas.
Given the low gas density in Eq.~(\ref{eq:nH0}), we consider the most 
optimistic scenario: If the gas around $r_m$ is able to cool to $10^4$~K 
(the equilibrium temperature between photo-ionization heating and 
collisional cooling) from $10^6$~K (the original stellar wind temperature)
while maintaining pressure equilibrium,
it may be undergo compression by a factor of 100. Thus the characteristic
hydrogen column density is
\be
N_H\sim 100n_H^{(0)} r_m\sim 3\times 10^{17}f_p^{1/6}\!
\left(f_{w1}^{-1}v_{w1}^{-7}M_{w1}^5\right)^{1/6}\!
B_{p1}^{1/3}\,{\rm cm}^{-2}.
\label{eq:NNH}\ee
Comparing this with Eq.~(\ref{eq:Ni}), we see that, under this optimistic
condition, sufficient metal column density may be attained when 
$n_i/n_H\go 5\times 10^{-5}$, which is of
the same order as the stellar abundance of Magnesium
\footnote{What is of interest here is the metal abundance of the stellar wind,
which is unknown. In the case of the Sun, it is known that the solar wind
metal abundance can be larger than the abundance at the solar atmosphere
(e.g., Bochsler \& Geiss 1989).}.
Obviously, a larger $\dot M_w$ would result in a higher gas column density.

It thus seems that the compressed gas at the magnetopause might
just be able to cause enhanced absorption before ingress.
The interaction of the stellar wind with planetary magnetospheres
is, however, much more complex than discussed above (e.g., 
Kivelson \& Russell 1995). Our estimate in Eq.~(\ref{eq:NNH})
suggests that such interaction may produce observable effects.
Further study on this subject in the context of hot Jupiters is worthwhile 
(cf.~Preusse et al.~2007; Ekenb\"ack et al.~2010).


\section{Discussion}

Motivated by the recent near-UV spectral observations of the WASP-12
planetary system (Fossati et al.~2010), which showed an earlier
ingress of the planet's transit compared to the optical continuum, we 
have studied two possible explanations for the ingress/egress
asymmetry. The first involves mass transfer through Roche lobe overflow,
which results in an elongated accretion stream in which gas flows 
from the L1 Lagrangian point toward the star. Our analysis of the geometric
and velocity structure of the stream shows that it may indeed provide
asymmetric obscuration of the star light during the planet transit.
In addition, it is also possible that an asymmetry in the accretion
disk, caused by the impact of the accretion stream, produces an
apparent earlier ingress.
Another possibility involves the magnetopause separating the
stellar wind and the planetary magnetosphere. Because of the planet's
orbital motion, the head magnetopause lies eastward relative to the
substellar point. We suggest that line absorption by the gas around the
magnetopause may also explain the asymmetric ingress/egress behavior,
although more theoretical works are needed to understand the property  
of the absorbing gas around the magnetopause. 

The two possibilities studied in this paper may be
distinguished by the fact that in the accretion stream, the gas falls away from 
the observers, producing redshifted absorption, while the flow around
the magnetosphere tends to move toward the observer, thus producing 
blueshifted absorption.

Finally, we note the observed behaviors reported by Fossati et al
are only marginally significant (at most $3\sigma$ effects).
Thus, more observations will be useful. Nevertheless, the physical
processes discussed in this paper are quite general and therefore may be
applicable in other close-in exoplanetary systems.


\medskip
We thank the participants of the morning coffee of
the KITP Exoplanets program (2010.1-2010.5) for lively discussions.
D.L. especially thanks Phil Arras, Doug Lin and Steve Lubow for useful
discussion during the final phase of our work, and P. Arras for 
his comments on an earlier draft of our paper. C.H. thanks Luca
Fossati for discussion of early-ingress observations. 
Part of this work was performed while the authors were in
residence at 
KITP, funded by the NSF Grant PHY05-51164.



\begin{thebibliography}{}

\bibitem[\protect\citeauthoryear{}{}{}]{}
Armitage, P.J. 2007, arXiv:astro-ph/0701485

\bibitem[\protect\citeauthoryear{}{}{}]{}
Bochsler, P., \& Geiss, J.~1989, in Solar System Plasma, eds. J.H. Waite et al.
(Washington DC: AGU), pp.~133-141

\bibitem[\protect\citeauthoryear{}{}{}]{}
Campo, Ch. et al. 2010, ApJ, submitted  (arXiv:1003.2763)

\bibitem[\protect\citeauthoryear{}{}{}]{}
Cranmer, S.R. 1998, ApJ, 508, 925

\bibitem[\protect\citeauthoryear{}{}{}]{}
Ekenb\"ack, A., et al.~2010, ApJ, 709, 670

\bibitem[\protect\citeauthoryear{}{}{}]{}
Fossati, L., et al.~2010, ApJ, 714, L222

\bibitem[\protect\citeauthoryear{}{}{}]{}
Garcia Munoz, A. 2007, Planet. Space Sci., 55, 1426

\bibitem[\protect\citeauthoryear{}{}{}]{}
Gu, P.-G., Lin, D.N.C., \& Bodenheimer, P.H. 2003, ApJ, 588, 509

\bibitem[\protect\citeauthoryear{}{}{}]{}
Hebb, L. et al.~2009, ApJ, 693, 1920-1928

\bibitem[\protect\citeauthoryear{}{}{}]{}
Hundhausen, A.J. 1995, in Introduction to Space Physics,
eds. M.G. Kivelson \& C.T. Russell (Cambridge Univ. Press), p.~91

\bibitem[\protect\citeauthoryear{}{}{}]{}
Husnoo, N., et al.~2010, MNRAS, submitted (arXiv:1004.1809)

\bibitem[\protect\citeauthoryear{}{}{}]{}
Kivelson, M.G., \& Russell, C.T. 1995, Introduction to Space Physics
(Cambridge Univ. Press)

\bibitem[\protect\citeauthoryear{}{}{}]{}
Koskinen, T.T., Yelle, R.V., Pavvas, P., \& Lewis, N.K. 2010,
arXiv:1004.1396

\bibitem[\protect\citeauthoryear{}{}{}]{}
Li, S-L., Miller, N., Lin, D. \& Fortney, J. 2010,
Nature, 463, 1054

\bibitem[\protect\citeauthoryear{}{}{}]{}
Lin, D.N.C., \& Papaloizou, J.C.B. 1993, in Protostars and Planets III, 
ed. E.H. Levy \& M.S. Matthews (Tucson: Univ. Arizona Press), 749

\bibitem[\protect\citeauthoryear{}{}{}]{}
Lubow, S.H., \& D'Angelo, G. 2006, ApJ, 641, 526

\bibitem[\protect\citeauthoryear{}{}{}]{}
Lubow, S.H., \& Shu, F.H. 1975, ApJ, 198, 383

\bibitem[\protect\citeauthoryear{}{}{}]{}
Lubow, S.H., \& Shu, F.H. 1976, ApJ, 207, L53

\bibitem[\protect\citeauthoryear{}{}{}]{}
Mestel, L., \& Spruit, H.C. 1987, MNRAS, 226, 57

\bibitem[\protect\citeauthoryear{}{}{}]{}
Murray-Clay, R.A., Chiang, E.I., \& Murray, N. 2009, ApJ, 693, 23

\bibitem[\protect\citeauthoryear{}{}{}]{}
Preusse, S., et al. 2007, Planetary \& Space Sciences, 55, 589

\bibitem[\protect\citeauthoryear{}{}{}]{}
Russell, C.T., \& Walker, R.J. 1995, in Introduction to Space Physics,
eds. M.G. Kivelson \& C.T. Russell (Cambridge Univ. Press), p.503

\bibitem[\protect\citeauthoryear{}{}{}]{}
Rybicki, G.B., \& Lightman, A.P. 1979, Radiative Processes in Astrophysics
(Wiley-VCH)

\bibitem[\protect\citeauthoryear{}{}{}]{}
Schneiter, E.M., et al.~2007, ApJ, 671, L57-L60

\bibitem[\protect\citeauthoryear{}{}{}]{}
Shu, F.H. 1992, Gas Dynamics (University Science Books)

\bibitem[\protect\citeauthoryear{}{}{}]{}
Shu, F.H., \& Lubow, S.H.~1981, ARAA, 19, 277

\bibitem[\protect\citeauthoryear{}{}{}]{}
Tian, F., et al. 2005, ApJ, 621, 1049

\bibitem[\protect\citeauthoryear{}{}{}]{}
Yelle, R.V. 2004, Icarus, 170, 167

\end{thebibliography}
\end{document}